\documentclass[manuscript,screen]{acmart}

\usepackage{makecell}
\usepackage{graphicx}
\AtBeginDocument{%
  }

\setcopyright{acmlicensed}
\copyrightyear{2024}
\acmYear{2024}
\acmDOI{XXXXXXX.XXXXXXX}

\begin{document}

\title{Vulnerabilities that arise from poor governance in Distributed Ledger Technologies}

\author{Aida Manzano Kharman}
\affiliation{%
  \institution{Imperial College London}
  \city{London}
  \country{United Kingdom}}
\email{aida.manzano-kharman17@imperial.ac.uk}
\orcid{0000-0002-5342-3037}
\author{William Sanders}
\email{william.sanders@iota.org}
\affiliation{%
  \institution{IOTA Foundation}
  \city{Berlin}
  \country{Germany}
}

\renewcommand{\shortauthors}{Manzano Kharman, Sanders}

\begin{abstract}
Distributed Ledger Technologies (DLTs) promise decentralization, transparency, and security, yet the reality often falls short due to fundamental governance flaws. Poorly designed governance frameworks leave these systems vulnerable to coercion, vote-buying, centralization of power, and malicious protocol exploits—threats that undermine the very principles of fairness and equity these technologies seek to uphold. This paper surveys the state of DLT governance, identifies critical vulnerabilities, and highlights the absence of universally accepted best practices for good governance. By bridging insights from cryptography, social choice theory, and e-voting systems, we not only present a comprehensive taxonomy of governance properties essential for safeguarding DLTs but also point to technical solutions that can deliver these properties in practice. This work underscores the urgent need for robust, transparent, and enforceable governance mechanisms. Ensuring good governance is not merely a technical necessity but a societal imperative to protect the public interest, maintain trust, and realize the transformative potential of DLTs for social good.
\end{abstract}

\ccsdesc[500]{Human-centered computing~Human computer interaction (HCI)}
\ccsdesc[500]{Security and privacy}

\keywords{Governance, Distributed Ledger Technologies, Privacy, Voting}

\received{20 February 2007}
\received[revised]{12 March 2009}
\received[accepted]{5 June 2009}

\maketitle

\section{Introduction} \label{intro}
Distributed Ledger Technologies (DLTs) have enabled a paradigm shift in how users interact and transact online. DLTs are comprised of users, that interact with the DLT network and receive some service from it, and service providers (miners and node operators) that maintain the DLT network by running its protocol and providing service to its users \cite{romero2018distributed}. DLTs do not exist in a single server, but rather run on a web of connected nodes, operated by the service providers, which communicate following a protocol \cite{sunyaev2020distributed}. 
A protocol is simply a set of instructions that nodes execute to enable the functioning of the DLT. DLTs are run through code, and maintainable code requires updates \cite{liu2020distributed}. DLTs must have a mechanism through which protocols are updated, bugs are patched and the service is improved. This component is crucial to ensure DLTs adapt, develop and improve in a sustainable manner, as well as providing a much needed protection against threats \cite{Accenture2019}. Without this ability to mutate, vulnerabilities cannot be patched. Since all nodes must agree to use the same protocol, a DLT must rely on a consensus algorithm to decide which protocol updates will take place and when. Governance is an all-encompassing umbrella term that captures this ability: the decision-making process to change a DLT protocol \cite{lehdonvirta2016governance}. 

Whilst some earlier work laid the foundations \cite{vishnumurthy2003karma}, \cite{dai1998b}, \cite{chaum1983blind}, \cite{back2002hashcash} it is the Bitcoin white paper \cite{nakamoto2008bitcoin} that is widely regarded to have kickstarted the Web 3 revolution. Satoshi Nakamoto proposes a peer-to-peer electronic cash system in the wake of the Great Recession of 2008. Blockchain was conceived as a reactionary and idealistic attempt to redistribute power and decentralise decision-making of financial service providers\footnote{Satoshi states in 2009: \emph{“The root problem with conventional currency is all the trust required to make it work. The central bank must be trusted not to debase the currency, but the history of fiat currencies is full of breaches of that trust. Banks must be trusted to hold our money and transfer it electronically, but they lend it in waves of credit bubbles with barely a fraction in reserve. We have to trust them with our privacy, trust them not to let identity thieves drain our accounts. Their massive overhead costs make micropayments impossible.”} \cite{Nakamoto2009}.}. 

Cryptocurrencies initially faced a shy uptake but exploded in value after 2017. At time of writing (November 2024), Bitcoin price is up by 24,801.30\% since it was created. Similarly, Ether increased by 1,985.73\%, Cardano by 929.34\% and Solana by 37,300\% \cite{forbes_crypto_prices}. 
Cryptocurrencies are here to stay: governments have not only begun to regulate these assets (examples include the US \cite{hughes2017cryptocurrency}, Japan \cite{arora2020cryptoasset} and the EU \cite{zetzsche2021markets}), but also own and invest in them as well \cite{mitchell_top_government_bitcoin_holders_2024}  (El Salvador \cite{gaikwad2021impact} and Brazil \cite{redman_brazil_bitcoin_reserve} actively investing, and other countries holding seized assets such as the US and UK are just some examples \cite{lee_gov_bitcoin_holdings}). 

Although Satoshi Nakamoto conceived Bitcoin as a decentralised, peer-to-peer network, the reality is that it and many other cryptocurrencies are centralised \cite{beikverdi2015trend} \cite{sai2021taxonomy}, manipulable \cite{dhawan2023new} \cite{barnes2018crypto} and can be undermined due to their governance protocols \cite{daian2019flash}, \cite{judmayer2019pay}. Governance is critical to delivering the promise of decentralisation, and without good governance, the vulnerabilities that may arise are dire. These issues must be addressed to preserve the original purpose of using cryptocurrencies for social good. Cryptocurrencies have become a matter of public interest: they are used by governments, their combined market worth is of \$3.71 Trillion \cite{forbes_crypto_prices} and the entry barrier to invest in them is very low, meaning many with low technical and financial literacy can and do participate and invest in these technologies \cite{bartlett_cryptoqueen_2019}. Whilst this is revolutionary, it is also highly dangerous: many victims have lost life's savings due to scams, hacks, heists and rug pulls with little to no compensation \cite{stanford2021crypto}. Governance safeguards are necessary to ensure transparency, accountability and prevent monopolisation and theft of funds. 

Our work addresses the governance dimension of DLTs that is increasingly intersecting with public policy, regulation, and societal well-being. Cryptocurrencies and distributed ledger technologies (DLTs) are no longer niche financial tools; they are evolving into systems that challenge traditional governance models and require new frameworks for oversight, trust-building, and public protection. By identifying vulnerabilities and proposing governance best practices, this work contributes to the interdisciplinary dialogue on how digital technologies transform governance structures, impact public institutions, and necessitate regulatory adaptations. Ultimately, we hope this research informs policy-making, aids regulators in understanding systemic risks, and helps ensure that these technological innovations serve, rather than exploit, the public interest.

\section{Overview}
This paper is organized as follows: Section \ref{intro} introduces the motivation for examining governance in Distributed Ledger Technologies (DLTs). Section \ref{related-work} reviews related work, identifying gaps in the current understanding and implementation of governance properties in DLTs and the vulnerabilities they face. Section \ref{methodology} provides a detailed overview of our methodology, Section \ref{components} introduces the governance components needed to outline our main findings. Subsequently, we describe vulnerabilities that may arise due to current governance methods in DLTs in Section \ref{vulnerabilities}, and we present a taxonomy of governance properties in Section \ref{properties}. The latter also includes possible technical solutions that may be implemented to deliver good governance properties. Section \ref{evaluation} evaluates a selection of DLT applications against our framework of governance properties, and identifies gaps and mutually exclusive properties. Finally, Section \ref{conclusion} concludes with a summary of our contributions and directions for future research.

\section{Related Work} \label{related-work}
A comprehensive systematization of knowledge on good governance properties can be found in \cite{kiayias2022sok} and \cite{liu2023systematic}. The authors in \cite{kiayias2022sok} focus on governance properties, whereas \cite{liu2023systematic} answer questions on what governance is, who is involved, when is it used and why. The work in \cite{kiayias2022sok} serves as a starting place for the work we present. Namely, we find their taxonomy of governance properties the most clear, tangible and concise thus far. However, they explicitly note that they do not provide specific techniques, algorithms or mechanisms to achieve good governance properties. 

The authors in \cite{khan2020blockchain} consider blockchain governance in the context of IT governance, and study different decisional outcomes using the Nash equilibrium. Other research efforts such as \cite{pelt2021defining}, \cite{beck2018governance} and \cite{de2018governance} aim to define governance and its components, although none delve deeply into tangible definitions of good governance properties. 

In terms of vulnerabilities that may arise as a consequence of poor governance in DLTs, \cite{schneider2022cryptoeconomics} claims that economic incentives alone do not suffice to enable good governance, and makes a case that well studied social choice and politics findings should be applied to governance in DLTs. The work in \cite{benedict2019challenges} outlines a number of challenges regarding DLT governance. Namely, it notes how poor governance models can cause oligarchies in DLTs, how incentives are crucial in ensuring accountability and notes how decision rights should be clearly defined in the governance protocol. The authors in \cite{de2016invisible} identify that governance in Bitcoin is not as decentralised and code-driven as expected, and uncovers that in fact the decision-making power is highly technocratic and determined by a small minority of code developers. None of the aforementioned address the formal security and privacy notions necessary in the governance protocol, or the vulnerabilities that arise in the absence of these.
We find a substantial overlap in the study of governance vulnerabilities in Decentralised Autonomous Organisations (DAOs) and DLTs. The work in \cite{austgen2023dao} and \cite{daian2018chain} outline how vote buying cartels may be automated as a consequence of poor governance models in DAOs. The authors in \cite{feichtinger2024sok} also highlight a number of governance related vulnerabilities in DAOs. Specifically regarding purchaseable voting power, malicious proposals and flash loans exploits, which the work in \cite{kharman2024perils} also discusses. A noteworthy body of research is carried out in \cite{tan2023open}, where the authors note the importance of classical security notions for e-voting should also hold in DAO governance mechanisms. We agree. However, they do not provide guidance on how to achieve these primitives. 
In summary, some recent works have emerged outlining vulnerabilities that arise as a consequence of poor governance models in DAOs, but less so regarding DLT governance. Some cautionary warnings have been raised regarding DLT governance, but we find no work that explores the suite of possible attacks that may arise in DLTs in the absence of standard good governance properties.

\section{Methodology} \label{methodology}
Our work employs a qualitative, analytical approach to identify vulnerabilities in Distributed Ledger Technologies (DLTs) arising from deficiencies in governance models. 

\subsection{Identification of Vulnerabilities}
From work previously carried out regarding DAO vulnerabilities in \cite{kharman2024perils}, we identified their governance protocols enable a number of weaknesses as a consequence of flaws in their governance protocols. Other researchers corroborate these findings: \cite{feichtinger2024sok}, \cite{austgen2023dao}, \cite{tan2023open}. We hypothesize that these vulnerabilities are also applicable to DLTs, since DAOs are implementations built on DLTs. We then proceeded to survey existing work following two criteria: governance related vulnerabilities, and governance properties of DLTs. 

\subsection{Identification of Governance Properties and Vulnerabilities}
\begin{figure}
    \centering
    \includegraphics[width=0.5\linewidth]{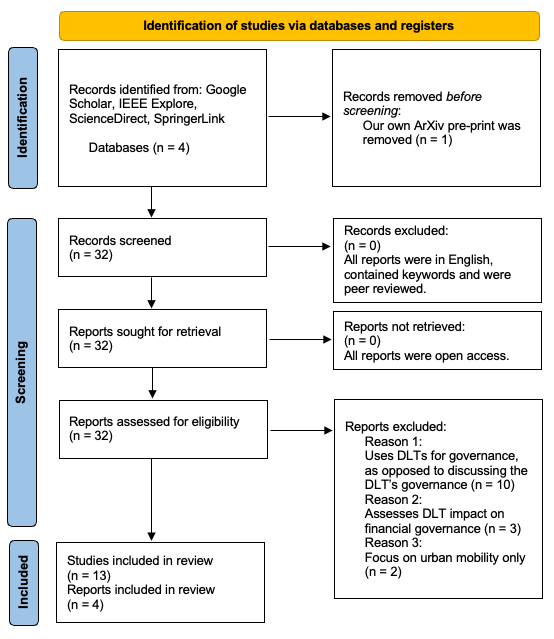}
    \caption{PRISMA flow diagram for systematic review of literature}
    \label{fig: prisma-review}
\end{figure}

A comprehensive review of existing literature and frameworks on governance in DLTs was conducted to identify key governance vulnerabilities and properties. 
This was done following the PRISMA methodology \cite{moher2010preferred}. We searched 4 databases using the keywords: \emph{DLT}, \emph{governance}, \emph{vulnerabilities}, \emph{attacks}, \emph{properties}. The outcome of this search is summarised in Figure \ref{fig: prisma-review}.
The number of works identified was limited, we therefore widened our search to include DAO governance, of which only the work of \cite{tan2023open}, \cite{feichtinger2024sok} and \cite{daian2018chain} were identified as relevant. 
Of the surveyed studies, only \cite{kiayias2022sok} focuses on governance \emph{properties} of DLTs, and \cite{gojka2021security} on \emph{vulnerabilities} of DLTs, although it does not explicitly discuss which of these vulnerabilities arise as a consequence of poor governance. The authors of \cite{tan2023open} point to some formal notions of good governance in the context of DAOs, and implementations for governance that attempt to deliver these. These, in conjunction with \cite{kiayias2022sok} serve as the starting point for our work regarding our contributions on governance properties, and \cite{daian2018chain} and \cite{gojka2021security} for our contributions on governance related vulnerabilities.

\subsection{Assessment of Governance Gaps}
Using the governance properties as a benchmark, an analysis of governance models in various DLT systems was performed. This step involved examining case studies, technical documentation, and academic studies to identify instances where these governance properties were insufficiently implemented or entirely absent. 

\subsection{Linking Vulnerabilities to Absent Governance Properties}
The identified governance failings were analysed to determine their impact on DLTs. Logical reasoning and case-based evidence were used to connect missing governance properties to specific vulnerabilities, such as vote buying, coercion, centralisation of power, incentives to heist the DLT and non-compliance with election outcomes.

\subsection{Theoretical Justification}
We then leveraged existing theories and frameworks in organizational governance, distributed systems, e-voting, cryptography and social choice theory to substantiate the connections between governance deficiencies and the emergence of vulnerabilities.

\color{black}
\section{Governance Components}\label{components}
In this work, we refer to \emph{governance} as the decision-making processes to enable the functioning and updating of a DLT protocol. The authors in \cite{zachariadis2019governance} similarly state that the term governance refers to `the frameworks and processes that oversee decision-making, control, and the evolution of the system. This encompasses how rules are established, enforced, and modified within the decentralized network'. The work in \cite{lehdonvirta2016governance} state that governance in DLTs is the `rule-making by the owners or participants of a system with the purpose of safeguarding their private interests'. Other approaches examine the intersection between blockchain governance and corporate governance \cite{yermack2017corporate} and borrow from the definitions of corporate governance \cite{davidson2024corporate}. For example, the work in \cite{yermack2017corporate} states: `Ultimately blockchains must rely on a governance process in which the users agree upon a set of requirements for the underlying software code to be changed, including provisions for dispute resolution, sanctions for violating the agreed upon rules, and procedures for enforcement of penalties'. Whilst there is no consensus on what is the definition of governance in the context of DLTs, most definitions overlap with the following critical functionalities: consensus, rules and decision-making.

Before proceeding with the outlined vulnerabilities, we establish definitions and nomenclature to aid with the explanation of governance related attacks. We present what we consider to be the minimum viable components to enable decision-making on-chain. These are borrowed from electronic voting literature, that also establish similar roles \cite{adida2008helios}, \cite{chaum2008scantegrity}, \cite{clarkson2008civitas}, \cite{ryan2009pret}.

\begin{itemize}
    \item The voters.
    \item The votes.
    \item The vote tallying algorithm.
    \item The election officials: these are participants in charge of tallying the votes, maintaining the voting platform and publishing results.
    \item The bulletin board: the digital record of cast ballots.
    \item Stakeholders: they may or may not be allowed to vote, but they are in some way affected by the outcome of an election. This includes participants in the DLT that are in charge of implementing the election outcome.
    \item The platform to vote: the server, website, chain or other online location where votes are cast.
\end{itemize}


\section{Vulnerabilities}\label{vulnerabilities}

In this section we outline the existing vulnerabilities that arise as a consequence of poor governance models. We highlight the intersection between DAO and DLT governance: both share similarities in their practical implementations. Indeed, this follows from the fact that many DAO projects are built on DLTs. 

\subsection{Exploits leveraging public votes}\label{secret ballots}
The right to a private vote is recognized as a human right as per Article 21.3 of the Universal Declaration of Human Rights \cite{assembly1948universal} in 1948. In the absence of it, voters may be coerced, bribed or sell their votes. Secret ballots are congruent to freedom, they ensure voters can express their will free of external influences and pressures. 
This is no different in DLT governance. 

\subsubsection{Voter Coercion}

When ballots are public, everyone can know how a voter voted. This provides verifiability: everyone can check that the election outcome is correct and that only those allowed to vote voted. However, this forgoes freedom. Free will cannot be protected when a voter faces the reality of not abiding with their coercer's threats \cite{lever2007mill}.
The coercer can easily check if their victim complied. Voters may chose to sell their votes even in the absence of coercion. 
In some cases, DLTs provide a unique set of conditions where the most beneficial strategy for some voters is to sell their vote. Other works have identified this issue in DAO governance \cite{daian2018chain}, \cite{austgen2023dao}. We highlight that this problem is prevalent in DLT governance too. 

Currently, most DLTs have informal channels of discussion where participants debate on whether the proposal will be implemented, and from what time on-wards it will be adopted if it has received enough support. These channels range from Slack, GitHub or Discord, to dedicated voting portals such as Snapshot and Tally, and often involve `rough consensus'.

We stress, at current time of writing, none of these solutions offer secret ballots. No existing solution can be shown to satisfy formal notions of ballot secrecy, because the ledger is public. Some have attempted to provide increased levels of privacy, but this privacy is short lived, and often compromised to achieve verifiability. An example is Snapshot's recently added feature of `shielded voting', wherein votes remain private only until the end of the election. Then, `when voting ends, all votes are revealed - as well as who voted for what' \footnote{\href{https://decrypt.co/105201/snapshot-adds-shielded-voting-daos-help-solve-voter-apathy}{Snapshot adds `Shielded Voting'}.}. This still enables voter coercion and vote selling. 

\subsubsection{Disincentive to vote against whales}
Both users and miners vote with their feet on a DLT. When a proposal arises for a protocol update, both must agree for it to be implemented. Users may favour a certain proposal, but miners can refuse to service transactions following it. Similarly, if miners favour a proposal that users disagree with, the users may simply stop sending transactions on that chain, and miners will earn no fees.
Let us consider a voter with much larger voting power than the rest a \textit{whale}. An example could be a user with a large share of tokens or a collective with high mining power (a large mining pool) in the DLT. When a whale votes for a certain proposal, other agents are incentivised to follow suit. Otherwise a fork\footnote{A fork is a divergence in a DLT regarding the accepted version of transactions \cite{wang2019corking}.} may arise, and eventually one of the two will outgrow the other and become the main chain. Miners processing transactions on the losing fork will not receive their reward. Consequently users' transactions will not get approved, and thus they too will eventually leave the fork. Because whales hold greater voting power, the fork they vote for will be the winning one with greater likelihood. Less influential voters will minimise their chance of losses by following the behaviour of whales, which is public, on-chain. This creates a herding effect, and a disincentive to vote true to one's beliefs, if these oppose those of a whale's.

For a voter in this situation, their only non-losing strategy is to abstain, but they are guaranteed a reward when they sell their vote. Given that votes are public, this sale can be verifiable: the vote buyer can check that they complied. Indeed, \cite{austgen2023dao} and \cite{daian2018chain} show how vote buying can be automated through a DAO cartel. Their prototype for vote buying code is designed to buy DAO votes, although there is no impediment to using it to buy DLT votes. 
\newline

\subsubsection{Voter Information Asymmetry}
When ballots are public, the first voter casting a vote has no information on what the election outcome will be, whereas the last voter to cast their vote knows the election outcome. Voters are unequal: the later you vote the more information you obtain regarding the election result. This allows more powerful voters to withhold their vote until they can guarantee that it is sufficient to sway the election result in their favour. 
 

Another simpler exploit consists of an attacker purchasing enough voting power to approve a malicious proposal in their favour. Public voting facilitates this attack: the attacker knows exactly how much voting power they need to acquire to overturn an election in their favour. 
These attacks have been witnessed in DAOs a number of times \footnote{\href{https://decrypt.co/92970/build-finance-dao-falls-to-governance-takeover}{Build DAO's hostile governance takeover attack, Feb 2022}.}, \footnote{\href{https://www.theverge.com/2022/4/18/23030754/beanstalk-cryptocurrency-hack-182-million-dao-voting}{Beanstalk cryptocurrency project robbed after hacker votes to send themselves \$182 million}.}, \footnote{\href{https://www.bloomberg.com/news/articles/2023-05-21/sanctioned-crypto-mixer-tornado-cash-hijacked-by-hackers}{Sanctioned Tornado Cash DAO governance heisted by hacker}.}, and are plausible in DLTs too. As shown in \cite{bitcoin-hashrate} and \cite{gencer2018decentralization}, miners' voting power (hash rate distribution) is highly centralized. In some cases, the voting power held by one agent can suffice to fork a network. In others, malicious users may leverage flash loans\footnote{Loans written in smart contracts that enable participants to quickly borrow funds without the need for collateral~\cite{eulerhack}.} to rapidly obtain sufficient tokens, submit a proposal to their benefit, cash out and repay the loan, whilst still making a profit. Miners are incentivised to service this proposal: they stand to gain high transaction fees. Indeed, the behaviour of whales is known to affect the price of a token \cite{manahov2022cryptocurrency}: they have the power to send large scale transactions that may incentivize miners to act against the consensus protocol \cite{liao2017incentivizing}, \cite{daian2020flash}.  
\newline

\subsection{Exploit leveraging non-binding elections} \label{binding-elections}

In a DLT context, we consider a binding election one in which the outcome of the election is enforced into the DLT. 

We note that, although informal voting channels may exist, these are in fact nothing other than a polling mechanism, because the only binding votes on a DLT protocol update are the transactions sent by users and approved by miners on-chain, following either one protocol or another.
All other off-chain so-called voting platforms are operating under the assumption that miners and users will behave on-chain in the same way that they expressed they would off-chain. The New York Agreement is a clear example of how off-chain voting is not binding and how this assumption may not always hold\footnote{\href{https://fullycrypto.com/what-was-the-bitcoin-new-york-agreement}{Bitcoin's New York Agreement} was a planned upgrade of Bitcoin to increase the block size from 1MB to 2MB. The update garnered significant community support but large Bitcoin mining pools refused to implement it \cite{bitcoin_independence_day}. This update would have lowered miner's transaction fees by improving transaction efficiency. The result was that Bitcoin was forked: Bitcoin maintained the same block size and Bitcoin Cash, the fork, adopted the increased block size \cite{webb2018fork}.} \cite{new-york-agreement}. Anyone can vote, but if the agents responsible for implementing the election outcome disagree with the outcome, they have no obligation to implement it, even if the election was fair, free and correct. This is a vulnerability, as stakeholders may subvert an election outcome by refusing to comply.
To prevent this, DLT governance protocols must be binding. To ensure this, elections must be carried out on-chain. On-chain voting brings a suite of challenges with it, which we outline in section \ref{on-chain voting}. 
\newline

\subsection{Exploits leveraging purchasable voting power}

\subsubsection{Plutocracy}
\emph{1-dollar-1-vote} mechanisms and any variant of these where voting power may be purchased will converge towards granting power to the most wealthy. Any governance mechanism that allows users to purchase voting power will inherently converge to plutocracy and centralise over time \cite{reijers2021now}. Whilst this issue is well known, DLTs continue to implement token based voting, for lack of a better alternative for a decentralised Sybil-protection\footnote{Sybil-protection mechanisms are those that prevent an attacker from creating multiple fake identities to gain unfair advantages in a network \cite{mohaisen2013sybil}.} system. 
DLT practitioners must compromise and choose which of the alternatives suits their application needs best:
permissionless voting does not rely on a central trusted service to prevent Sybil attacks, and is often implemented by providing voting rights proportional to wealth or computational power. However, it is known to centralise voting power in the hands of the wealthy.
Permissioned voting introduces reliance on a centralised trusted authority to issue voting rights, or requires a practical and functional decentralised identity management solution. This permissioned alternative is often implemented through reputation based voting schemes, soulbound tokens, NFTs, or a Know Your Customer solution, checking the validity of participants' identity.

\subsubsection{On-Chain Voting}\label{on-chain voting}
One of many concerns of on-chain voting is the transaction fees associated with a vote. No DLT currently offers zero transaction fees, thus, when miners or users wish to vote, they must pay to do so. This further contributes to exacerbate inequalities as only the wealthier can afford to do so. These fees often soar unpredictably \cite{JAIN2023104507}, and may also discriminate between voters as an election is occurring. The authors in \cite{park2021going} have extensively documented many of the concerns of blockchain voting.
Another concern pertaining to on-chain voting is the fact that censorship resistance cannot be assured \cite{wahrstatter2023blockchain}: miners may simply refuse to approve blocks of certain users. Conversely, malicious agents can spam the network with empty blocks, thus delaying the approval of the user's transactions. The feasibility of the latter highly depends on the access control mechanism of the DLT and the congestion control mechanism. 
Finally, current on-chain voting solutions do not offer ballot secrecy guarantees. The consequences arising from this are outlined in section \ref{secret ballots}.

It can therefore be observed that another compromise arises:
binding elections are offered by on-chain voting, however, this means censorship resistance is not guaranteed, and participants must pay to vote. 
\newline

\subsection{Other vulnerabilities}
There are other security concerns that arise when a combination of desirable properties are absent in the governance solution of the DLT. 

\subsubsection{Bribe Automation} 
This exploit is a consequence of the following properties being absent: ballot secrecy, non-purchaseable voting power and fee-less voting.
In DLTs where mining is competitive, if one mining pool gains a transaction, the other do not, so we can model the mining pools to be explicitly distinct from each other, and the smallest mining pool is of size 1 agent. 
Public votes make it evident to see how a mining pool voted by simply looking up the on-chain transactions. Less influential miners should either align their vote to those of the largest mining pools, or risk having their transactions end up in a fork and not receive the reward for the work they performed to include it. It is therefore easy for a briber to identify miners of less influence and bribe them to `vote' for their desired protocol. The small sized mining pools will make a profit, and the briber can identify how many miners they must bribe to succeed in implementing their desired protocol. As the authors in \cite{austgen2023dao} showed, paradoxically, the more decentralised the network, the easier it is to create a systematic bribing cartel. Small sized miners are cheaper to bribe than large mining pools. The authors even provide a protocol to automate this vote buying in DAOs. This can be re-purposed to buy a miner's vote in a DLT. The bribe can be paid out as soon as the smart contract detects a transaction following the desired protocol of the briber.

\subsubsection{Submitting Malicious Protocol Updates}
Malicious protocol updates are not infrequent, and not only in DAOs or DLTs\footnote{\href{https://thehackernews.com/2024/03/urgent-secret-backdoor-found-in-xz.html}{Backdoor found in Library update, used by major Linux Distributions}.}. Some go undetected due to low participation rates in governance, in others, the protocol update is forcibly approved by a whale, and in other occasions, the malicious protocol is hidden or not evidently detectable by the community. The consequences of this can be dire: double spending attacks, delaying governance outcomes, creating new tokens or loss of funds are amongst the many attack vectors possible. 

It is not feasible to characterise all the ways in which malicious protocol updates may be submitted. Governance is only one of the components necessary in the DLT to prevent these exploits, but a robust governance protocol does not guarantee protection against malicious protocol updates. Access control mechanisms, security and trust assumptions are other components that must be taken into consideration. We encourage DLT design practitioners to make choices in their governance protocol that will alleviate the risk of this exploit. Namely, removing incentives to submit malicious proposals, or to comply with bribes to implement them. 


\color{black}
\section{Governance Properties}\label{properties}
\subsection{Participation}
First we begin by considering who may participate in the decision making process of DLTs. These actors can be classified into the following set of agents: users, service providers, proposers and decision-makers. We introduce some terminology for ease of exposition.  

\begin{definition}[User]
    A user is a participant of the DLT network that interacts with the DLT to receive a service from it.
\end{definition}

\begin{definition}[Service providers]
    A service provider runs a set of agreed upon instructions (protocol) in their own machine that enables the functioning of the DLT.
\end{definition}

\begin{definition}[Decision-makers]
    A decision-maker is an agent that has the power to vote on a new proposal.
\end{definition}

\begin{definition}[Proposer]
    A proposer is an agent that can put forth a protocol update which will be subject to vote.
\end{definition}

Each of these sets of agents may overlap, service providers may also be proposers, users may also be decision-makers and so forth. The intersection of these sets is application specific to each DLT. 
There are two questions that must be addressed with regards to participation: suffrage and distribution of power. The former addresses who is allowed to vote, and the latter, how power is distributed amongst the voters.

We advocate for equal suffrage. This property means everyone eligible to vote has the right to cast a vote, and that everyone's vote carries the same weight \cite{shaw1914equal}. This property removes the possibility of weighted votes, thereby enforcing equal voting power distribution. 

This property may be too strict in some contexts. Some DLTs may wish to grant voting rights proportional to reputation or trustworthiness. However, we strongly discourage practitioners from distributing voting power proportionally to wealth.

In summary, two important properties of governance participation are:

\begin{enumerate}
    \item[\textbf{P1}] Equal suffrage.
    \item[\textbf{P2}] Non-purchasable voting power distribution.
\end{enumerate}

Note that \textbf{P2} is a weaker notion of \textbf{P1}. \textbf{P1} states each voter should receive one vote and all votes should be of equal weight. \textbf{P2} allows for votes of different weights, as long as the weight cannot be purchased.  \newline

\begin{figure}
\includegraphics[width=0.5\textwidth]{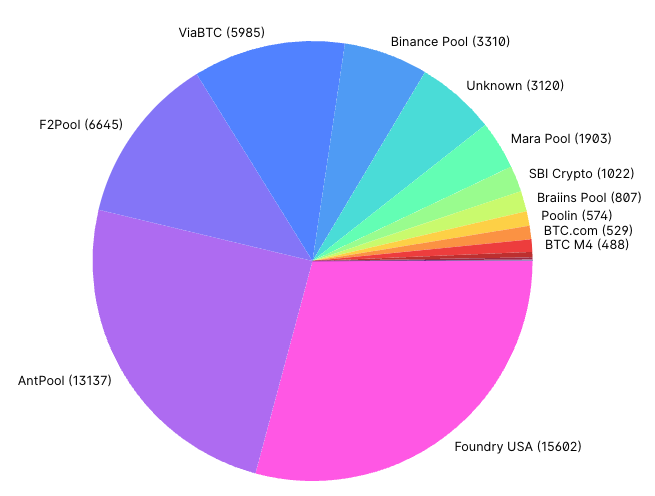}
\caption{Bitcoin's hashrate distribution for the year of 2024. Source: \href{https://www.blockchain.com/explorer/charts/pools}{Blockchain.com}} \label{blockchain hashrate}
\end{figure}
\begin{figure}
\includegraphics[width=0.6\textwidth]{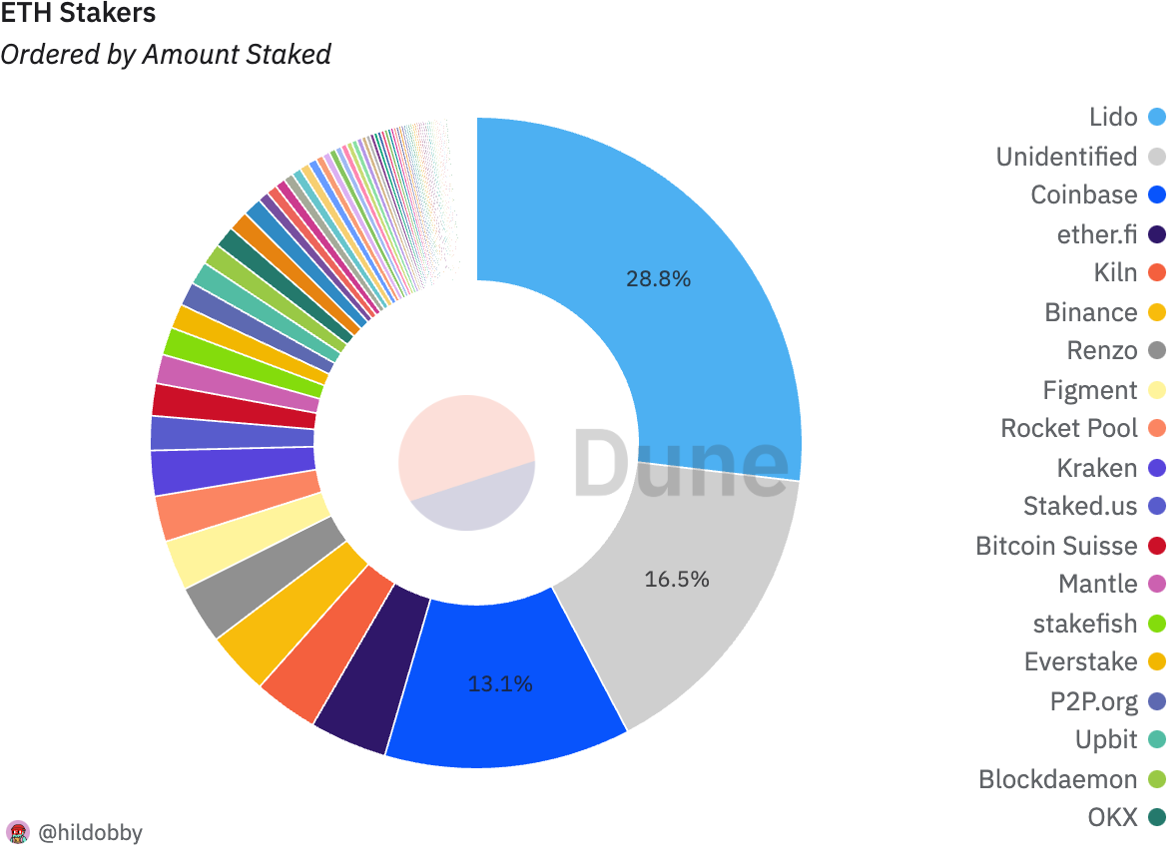}
\caption{Ethereum's staking pool distribution for the first 6 months of 2024. Source: \href{https://dune.com/hildobby/eth2-staking}{Dune.com}} \label{ethereum distribution}
\end{figure}

\noindent\textbf{Possible implementations:} Property \textbf{P1} advocates for a one-person-one-vote mechanism to distribute voting power. This can be easily achieved in permissioned DLTs, since each participant's identity is known. 
In permissionless DLTs, this is more challenging to achieve if a decentralised structure is to be retained\footnote{It can be argued that some permissionless DLTs are already not truly decentralised if the majority of voting power is held by a small number of entities and participation rights are assigned proportional to wealth. See Figure \ref{blockchain hashrate} for a distribution of hashrate in Bitcoin (regulated through Proof-of-Work) and Figure \ref{ethereum distribution} for the staking pool distribution of Ethereum (regulated through Proof-of-Stake).}. 

This is because it is non-trivial to distribute unique identity credentials in a decentralised manner, and in a way in which a single user cannot easily obtain multiple identity credentials. 
Indeed, an ongoing research effort is dedicated to answering this question \cite{dunphy2018decentralizing}, \cite{goodell2019decentralized}.  
The Proof of Humanity project aims to address this challenge. 
It is a peer-to-peer protocol that issues unique identity credentials to its participants\cite{PoH21}. 
They each create an identity, and it is then approved by other validated participants in the protocol. 
It offers the benefit that it is a more decentralised solution (it distributes trust), but participants must wait for others to vouch for their profile, and it is built on Ethereum so it is only available on their ecosystem. Bridges are protocols that allow for the transfer of data or transactions across different blockchains, and may be leveraged in this case to enable Proof of Humanity on other compatible blockchains \cite{bridges}. 

Semaphore is an Ethereum based zero-knowledge protocol that allows members to cast a vote without revealing their identity, whilst demonstrating they belong to a group of valid voters \cite{semaphore}.
In this manner, voters can only participate if they prove in zero-knowledge that they own a valid, unique voter credential, and can vote without having to reveal it.
It is possible to assign one person one vote in permissionless DLTs by combining a protocol such as Proof of Humanity with Semaphore. Proof of Humanity ensures only approved (real) individuals receive a unique identity credential. Then, voters can prove they own such a credential in zero-knowledge, meaning they do not need to publicly disclose their credential, through Semaphore. The work in \cite{10664257} proposes combining these two protocols in the context of resolving disputes is Web 3 platforms. 
We note both Proof of Humanity and Semaphore are only compatible with Ethereum. We encourage DLT engineers and researchers to prioritise delivering these functionalities across varying DLT architectures.

In summary, one-person-one-vote is a fair and democratic way of allocating voting rights but it will come at the expense of re-introducing some form of centralisation in permissionless DLTs to regulate who is considered a valid voter, and ensuring they only receive one unique identity credential.


\subsection{Incentives}
Many DLT protocols assume their governance system is secure due to the financial incentives offered to their participants. Unfortunately, in a number of cases these governance protocols are not clearly defined, documentation is sparse, outdated or spread across multiple sources \cite{ethereum-governance}, \cite{iota-gov}, \cite{iota-gov2}. Some protocols are so complex it is non-trivial to outline a clear mathematical model of how all of the incentives interact and what the net best strategy for an agent would be \cite{polkadot_governance}. As a result, undesirable scenarios may arise whereby agents are inadvertently incentivised to undermine the network \cite{zook2018new}.
There are two crucial incentives that must be clearly defined and identifiable in the governance protocol:
\begin{enumerate}
    \item[\textbf{I1}] Incentive to vote.
    \item[\textbf{I2}] Incentive to make useful and beneficial proposals.
\end{enumerate}

Property \textbf{I1} encapsulates whether the voter has an incentive to participate honestly in the voting protocol, and \textbf{I2} ensures there is no incentive to make undermining proposals. 

We then consider two more necessary properties:
\begin{enumerate}
    \item[\textbf{I3}] Distributed liability: risk is appropriately distributed amongst the governance participants. 
    \item[\textbf{I4}] Binding outcome: the outcome of the election must be implemented, provided the outcome is valid.
\end{enumerate}

\textbf{I3} is necessary to encourage the implementation of valid election outcomes. If service providers have no liability for not complying, \textbf{I4} is hard to enforce. Service providers may simply choose to not implement the outcome, without incurring on any cost for doing so. DLT practitioners must select an appropriate function to distribute risk, which is context specific to their own application. We highlight, it does not suffice with distributing liability amongst agents: it must be clear and traceable how this liability is distributed as well. 
Election outcomes must also be binding, otherwise the governance mechanism is simply a polling or referendum mechanism. 

\noindent\textbf{Possible implementations:} Some DLTs implement the property of distributed liability (\textbf{I3}) by introducing penalties for non-compliance (also known as slashing \cite{he2023don}), which also creates an incentive to participate honestly in the governance protocol. Implementing properties \textbf{I1} and \textbf{I2} is straightforward: direct rewards can be granted to participants upon voting or if the proposals they made receive sufficient support and are correctly implemented. Many cryptocurrencies already provide these rewards \cite{cardano_governance, polkadot_governance, solana_governance}. To implement \textbf{I4}, elections must be held on-chain, and those with the power to implement the election outcome must be able to participate in the election. The challenge lies in being able to characterise how all these incentives interact with each other, and whether other incentives in the cryptocurrency negatively interfere with them. For example, on-chain voting will introduce fees to cast a vote, and this may directly interfere with a financial incentive offered to voters for participating. 

\subsection{Voting Mechanism} 
Social choice theory has long studied the different properties that vote aggregation algorithms offer. Arrow famously demonstrated that no ordinal\footnote{Ordinal voting schemes only capture preference, whereas cardinal schemes reflect how much more one candidate is preferred over another.} voting scheme can simultaneously satisfy three seemingly simple and desirable properties: 
\begin{enumerate}
    \item[\textbf{V1}] Independence of Irrelevant Alternatives: If A is preferred over B, then introducing an irrelevant option C should not affect the outcome. 
    \item[\textbf{V2}] Pareto efficient: if the majority of voters prefer A over B, the voting system should output A over B.
    \item[\textbf{V3}] Non-dictatorial: there is no one voter that can single-handedly determine the election outcome.
    
\end{enumerate} \cite{arrow1950difficulty}. The Gibbard–Satterthwaite theorem further outlines that no ordinal voting scheme is simultaneously non-dictatorial and robust against tactical voting \cite{satterthwaite1975strategy}. In other words, that it does not satisfy \textbf{V3} and \textbf{V4} where \textbf{V4} is defined as:
\begin{enumerate}
\item[\textbf{V4}] Strategy-proofness: The best way for a voter to achieve their preferred outcome is to vote according to said preference.
\end{enumerate}

This does not bode well for ordinal voting schemes, but none of these results apply to cardinal voting schemes. A caveat with cardinal voting schemes is that it is important to find a meaningful function to allocate value to preferences. One example is Quadratic Voting \cite{lalley2016quadratic}, whereby each extra vote for a candidate is quadratically more expensive for a voter. Another class of voting mechanisms are those that are probabilistic\footnote{A probabilistic voting scheme returns an outcome depending on a certain probability. The same election run multiple times may not always return the same winner \cite{burden1997deterministic}.}. An example is the serial random dictator model or Maximum Entropy Voting \cite{sewell2009probabilistic}. Probabilistic voting mechanisms offer desirable properties such as \textbf{V1}, \textbf{V2} and \textbf{V4} (or slightly weaker notions of the latter two), and in some instances even \textbf{V3} \cite{sewell2009probabilistic}, \cite{bade2015serial}, \cite{kosheleva2021ranking}, \cite{bade2020random}. They also completely sidestep any of the theorems outlined by Arrow, Gibbard \cite{gibbard1973manipulation} and Gibbard–Satterthwaite.

\noindent\textbf{Possible implementations:} 
We strongly advocate for cardinal (provided a meaningful utility function to allocate preference intensity exists) or probabilistic schemes. These have been shown to satisfy the highest number of desirable properties by researchers in the field of social choice. 


\subsection{Security}
In this section we present a non-exhaustive survey of formal notions of security for online voting schemes, and existing implementations that satisfy these. We highlight that there is no academic consensus as to what the formal, cryptographic definition of these notions should be. DLT practitioners should select the primitives that offer the security level necessary or desired for their application.

Let us proceed by outlining the core notions of secure voting schemes. They must ensure the following properties are satisfied \cite{jardi2012study}, \cite{neumann1993security}, \cite{sabina-security}:
\begin{itemize}
    \item[\textbf{S1}] Ballot Secrecy: it should not be possible to determine how a voter voted from their ballot. 
    \item[\textbf{S2}] Verifiability: voters should be able to assert that the election outcome reflects the will of the voters. 
    \item[\textbf{S3}] Coercion resistance: a voter can not comply with the instructions of an adversary on how to vote, and this behaviour cannot be detected by the adversary. 

\end{itemize}

\noindent\textbf{Ballot Secrecy:} The quest to formally define \textbf{S1} is non-trivial. A recent survey of formal definitions of Ballot Secrecy is available in \cite{smyth2021ballot}. In it, the author also presents a formal definition of ballot secrecy. It is the state-of-the-art and it accounts for an adversary that has the power to intercept ballots as they are cast. In the same work, Smyth also proves that a variant of Helios voting scheme, namely a variant of Helios 3.1.4 satisfies ballot secrecy. This variant of Helios is known as Helios'16 and can be found in \cite{smyth2017election}. \\
\noindent\textbf{Possible implementations:} \label{bs-schemes} A number of voting scheme implementations exist that satisfy notions of ballot secrecy. Well studied proposals include Helios\cite{adida2008helios}, JCJ \cite{juels2005coercion}, Civtas \cite{clarkson2008civitas} and Athena \cite{smyth2019athena}. All of these protocols offer encrypted ballots, but not all provably satisfy the same notions of ballot secrecy, some have weaker security (earlier versions of Helios), others stronger, (JCJ and Civtas) but pay a price for it with an increased computational complexity. We note that none of these schemes have been implemented on-chain, although there are numerous voting smart contracts available. Whilst other on-chain voting schemes exist, most do not provide encrypted ballots. We find VoteCoin \cite{VoteCoin} and MACI \cite{maci} are the only on-chain voting schemes that provide encrypted ballots. Unfortunately, VoteCoin provides encrypted ballots only while the election is occurring. This forgoes secrecy once the election is over, enabling coercers to check if their victim complied. The same can be said about the work in \cite{kovst2019blockchain}. In MACI, votes remain encrypted, but the coordinator tasked with tallying individually decrypts ballots. If these are intercepted or if the coordinator is not trustworthy, votes may be easily revealed. Note that other protocols exist that provide anonymous voting, whereby the vote is public but the identity of the voter is hidden. Examples include Snapshot \cite{snapshot} and Vocdoni \cite{vocdoni2021}. This is \textbf{not} congruent to secret ballots and should not be considered an appropriate security measure especially in DLT implementations, where identities are at best pseudonymous \cite{androulaki2013evaluating}. 

In summary, DLT practitioners have a number of options to ensure their governance protocol satisfies notions of ballot secrecy. They may run off-the-shelf protocols such as Helios'16, Civtas, Athena or others previously mentioned. These can be run off-chain, and the results may be notarised on-chain leveraging zero-knowledge proofs. These can be used to demonstrate the election result was correctly computed. Alternatively, if on-chain voting is essential, practitioners may settle for options such as MACI or VoteCoin, which offer encrypted ballots, but these options are insecure and neither meet formal notions of ballot secrecy. 


\noindent\textbf{Verifiability:} An informal notion of \textbf{S2} in electronic voting schemes is end-to-end verifiability. If this is satisfied, a voter should be able to verify their ballot was:
\begin{itemize}
    \item cast as intended,
    \item recorded as cast,
    \item tallied as recorded.
\end{itemize} \cite{benaloh2015end}.
Another informal notion of verifiability is universal and individual verifiability. The authors in \cite{kremer2010election} state verfiability can be decomposed into:
\begin{itemize}
    \item Individual verifiability: voters can check that their own ballots are recorded \footnote{The property of individual verifiability is closely related to censorship resistance, but it is a weaker notion than the latter. Individual verifiability allows a voter to detect if their vote has been suppressed, whereas censorship resistance demands that votes cannot be suppressed in the first place.}.
    \item Universal verifiability: anyone can check that the tally of recorded ballots is computed properly.
    \item Eligibility verifiability: anyone can check that each tallied vote was cast by an authorized voter.
\end{itemize}
The work in \cite{smyth2015computational} then provides new formal cryptographic definitions for the three aforementioned. 

Whilst the informal notion of verifiability is rather intuitive, formalising such notion has been an arduous endeavour. For a comprehensive review of the literature and a proposed formal notion of verifiability, the reader may refer to the work in \cite{sok-verifiability}. 

\textbf{Possible implementations:} regarding available implementations, all e-voting schemes satisfy notions of verifiability (some weaker than others). It becomes a matter of deciding the level of security and the trust assumptions one can make in their system, to pick the right verifiability formalisation. Achieving verifiability alone is trivial, as can be seen from most blockchain voting implementations: simply publicly revealing votes yields verifiable voting schemes. The challenge lies in achieving both verifiability and secrecy, without sacrificing one in favour of the other. We stress: practitioners do not need to compromise, achieving both is indeed possible, and is in fact necessary. The possible implementations of ballot secrecy outlined earlier in \ref{bs-schemes} provide verifiability as well as ballot secrecy (granted that each adhere to different formal notions of both ballot secrecy and verifiability).

\noindent\textbf{Coercion Resistance:}
The work in \cite{juels2005coercion} introduced a paradigm shift in the concept of coercion-resistance. 
Prior to it, Benaloh and Tunistra introduce the notion of receipt-freeness as means to ensure voters cannot prove to a thrid party how they voted \cite{benaloh1994receipt}. 
However, the authors in \cite{juels2005coercion} consider a much more powerful adversary, one that may `demand of coerced voters that they vote in a particular manner, abstain from voting, or even disclose their secret keys'. 
They provide a formal notion of coercion resistance, and an election protocol that provably satisfies said notion. Note that this definition of coercion resistance partly captures censorship resistance\footnote{The property of censorship resistance means it should not be possible to suppress a valid voter's ballot \cite{wahrstatter2024blockchain}.}, as the adversary may demand of the voter to abstain. However, it does not consider an adversary that may suppress a ballot from the tallying phase. 
A recent survey of coercion resistant voting schemes can be found in \cite{battagliola2024algebraic} in section 6.5.1, and their respective time complexities is shown in Table 6.2. None of these schemes are implemented on-chain. The authors in \cite{spadafora2020coercion} propose a coercion resistant voting protocol using blockchain to cast votes. They also propose a security notion of Vote Indistinguishability (VI), which is congruent to secret ballots, and show their voting protocol is VI secure. 

\textbf{Possible implementations:} for on-chain voting, DLT practitioners may implement \cite{spadafora2020coercion}. Alternatively, one can select any of the aforementioned e-voting schemes that are coercion resistant (for example \cite{clarkson2008civitas}), run them off-chain and notarise the result on-chain leveraging zero-knowledge proofs. Finally, a number of other coercion-resistant schemes have been proposed but not implemented. These may be implemented on-chain if the DLT of choice allows for smart contracts in layer 1, or similarly, implement them off-chain and notarise the result on-chain.

\section{Evaluation}\label{evaluation}
We proceed by selecting a diverse range of cryptocurrencies to evaluate varying types of governance models, DLT architectures and functionalities. Bitcoin and Ethereum serve as foundational examples of blockchain systems with informal or off-chain governance processes \cite{bitcoin_governance}, \cite{ethereum_governance}.
Cardano is an interesting case-study, as its participants just recently approved a constitution to formalise its governance process\footnote{\href{https://cardanoconvention.com/}{Cardano Constitutional Convention 2024}.}. IOTA is an example of a Directed Acyclic Graph DLT, as opposed to a blockchain architecture \cite{popov2018tangle}. Solana was selected due to its alternative consensus mechanism, Proof of History, as well as its high transaction per second rates and low transaction fees \cite{yakovenko2018solana}. Avalanche is selected due to its consensus mechanism, the integration of subnetworks\footnote{The subnetwork architecture is one in which participants validate different subsets of blockchains in the network \cite{avalanche_whitepaper}.}, and the inclusion of multiple built-in blockchains \cite{avalanche_consensus_whitepaper}. 
Terra Luna was included as a case-study of a cryptocurrency that collapsed partly due to its governance structure, which failed to implement effective safeguards or timely interventions to address vulnerabilities in its algorithmic stable-coin model \cite{liu2023anatomy}, \cite{briola2023anatomy}.

\begin{table}[ht]
\centering
\resizebox{\textwidth}{!}{%
\begin{tabular}{|c|c|c|c|c|c|c|c|c|}
\hline
\textbf{Property} & \textbf{Bitcoin} & \textbf{Ethereum} & \textbf{IOTA} & \textbf{Polkadot} & \textbf{Cardano} & \textbf{Solana} & \textbf{Avalanche} & \textbf{Terra Luna} \\ \Xhline{3\arrayrulewidth}
Equal suffrage & No & No & No & No & No & No & No & No \\ \hline
Non-purchaseable voting power & No & No & No & No & No & No & No & No \\ \Xhline{3\arrayrulewidth}
Incentive to vote & None & Social & Social & Monetary & Monetary & Monetary & Monetary & Monetary \\ \hline
Incentive to make beneficial proposals & None & Social & Social & Monetary & Monetary & Monetary & Monetary & Monetary \\ \hline
Distributed liability & Yes (Miners) & Yes (Miners, Validators) & Yes (Users) & Yes (Users, Validators) & Yes (Users, Validators) & Yes (Validators) & Yes (Validators) & Yes (Both, Organising Foundation) \\ \hline
Binding election outcome & No & No & Yes & Yes & Yes & Yes & Yes & Yes \\ \Xhline{3\arrayrulewidth}
Independence of irrelevant alternatives & No & No & No & No & No & No & No & No \\ \hline
Pareto efficient & Yes & Yes & Yes & Yes & Yes & Yes & Yes & Yes \\ \hline
Non-dictatorial & Yes & Yes & Yes & Yes & Yes & Yes & Yes & Yes \\ \hline
Strategy-proofness & No & No & No & No & No & No & No & No \\ \Xhline{3\arrayrulewidth}
Ballot secrecy & No & No & No & No & No & No & No & No \\ \hline
Verifiability & Yes & Yes & Yes & Yes & Yes & Yes & Yes & Yes \\ \hline
Coercion resistance & No & No & No & No & No & No & No & No \\ \hline
Censorship resistance & Yes & Yes & Yes & Yes & Yes & Yes & Yes & Partially \\ \Xhline{3\arrayrulewidth}
\end{tabular}%
}
\caption{Governance properties satisfied in varying cryptocurrencies}
\label{tab:example_table}
\end{table}

For each of the selected cryptocurrencies, we survey their governance documentation and review the findings regarding each governance property below. The sources used for this section are the governance documentation pages of each of the cryptocurrencies: Bitcoin \cite{bitcoin_governance, nakamoto2008bitcoin}, Ethereum \cite{ethereum_governance, ethereum-governance}, IOTA \cite{iota_governance, iota-gov, iota-gov2},
Polkadot \cite{polkadot_governance}, Cardano \cite{cardano_governance}, Solana \cite{solana_governance}, Avalanche \cite{avalanche_governance, avalanche_consensus_whitepaper} and Terra Luna \cite{terra_luna_governance}.

\subsection{Participation}
Regarding suffrage, whilst none of the selected cryptocurrencies satisfy equal suffrage, token-based systems still allow all token holders to participate in governance. In that sense, suffrage is not completely denied to any eligible participant. The inequality arises in \emph{voting power}, not in the right to vote. All selected cryptocurrencies use some variant of token-based participation except Bitcoin, that grants participation rights proportionally to computational power.  
None deliver non-purchasable voting power: all selected examples use either Proof of Work, Proof of Stake or a variant of either to achieve consensus. In both of these systems, users may acquire greater power in the network by either purchasing more computational power or investing more wealth in the DLT network.

\subsection{Incentives}
Incentives to vote are delivered by Polkadot, Cardano, Solana, Avalanche and Terra Luna through direct financial compensation to the voters, whereas the incentive in IOTA is indirect. Stakeholders are motivated by the ecosystem's success. Ethereum does not offer direct recompense to voters, but applications built on the Ethereum ecosystem often do, such as DAOs. Bitcoin offers no financial reward or alternative incentives for participants to vote in the decision-making processes.

Similarly, Bitcoin offers no rewards or incentives to make beneficial improvement proposals. Bitcoin lacks a formal governance process for proposals. Changes are made via Bitcoin Improvement Proposals (BIPs), which require volunteer effort without direct economic incentives for proposers \cite{bitcoin_governance}. No built-in mechanism rewards making beneficial proposals, so participants rely on altruism or indirect benefits, such as network improvement. Ethereum also lacks direct incentives to make proposals, which are submitted as Ethereum Improvement Proposals (EIPs) \cite{ethereum_governance}. IOTA’s governance includes mechanisms for proposing changes, but there are no explicit financial incentives for proposers \cite{iota_governance}. Participants may be motivated indirectly by the chance to improve the protocol or gain reputation within the ecosystem. The remaining cryptocurrencies all have mechanisms to fund winning proposals, with Terra Luna, Cardano and Polkadot providing said funding from their own treasury \cite{terra_luna_governance, cardano_governance, polkadot_governance}. 

Bitcoin and Ethereum rely on voluntary adoption and off-chain agreement, making their governance outcomes non-binding. The rest of selected cryptocurrencies all implement on-chain voting, thereby making the outcome automatically binding. 

Regarding liability, Bitcoin miners are responsible for consensus and validation.
In Ethereum, miners were responsible when Ethereum used Proof-of-Work for consensus, and validators\footnote{Validators are the service providers in DLTs that use Proof-of-Stake for consensus.} are responsible since Ethereum shifted to Proof-of-Stake consensus. 
Participants indirectly manage the DLT in IOTA, since they can be both users and service providers at the same time. Validators and users share liability for governance and consensus in Polkadot, Cardano, Solana and Avalanche.
However, Terra Luna operated differently prior to its collapse: both users and the Organising Foundation (Terraform Labs) assumed liability in the governance process.
\subsection{Voting Mechanism}
The table below summarises the voting aggregation algorithms used in each cryptocurrency. Most use a token-based variant of plurality voting (also known as first-past-the-post or majority voting). Majority voting is known to satisfy Pareto Efficiency and is non-dictatorial, but it is not independent of irrelevant alternatives and therefore is not strategy proof \cite{fptp-iia}. The exceptions are Bitcoin and Ethereum, neither of which use on-chain voting. Bitcoin nonetheless satisfies the same properties because it also utilises a majority voting scheme, although the 95\% threshold is likely to reduce the influence of irrelevant alternatives in practise, compared to the 51\% threshold. We are unable to characterise the properties that Ethereum's voting scheme satisfies because there are no clear quorum or thresholds defined for adoption. 

\begin{table}[ht]
\centering
\begin{tabular}{|c|p{12cm}|}
\hline
\textbf{Cryptocurrency} & \textbf{Threshold for Proposal Approval} \\ \hline
\textbf{Bitcoin} & 95\% of miner signalling. \\ \hline
\textbf{Ethereum} & `Rough consensus'. \\ \hline
\textbf{Solana} & 50\%+1 of staked SOL (quorum rules may apply). \\ \hline
\textbf{IOTA} & 50\%+1 of Mana (majority-based). \\ \hline
\textbf{Cardano} & Planned: 50\%+1 ADA + quorum (higher for major changes). \\ \hline
\textbf{Polkadot} & Conviction-weighted majority, often >66\% for critical votes. \\ \hline
\textbf{Avalanche} & 50\%+1 AVAX (quorum-dependent). \\ \hline
\textbf{Terra Luna} & 40\% quorum; 50\%+1 approval threshold. \\ \hline
\end{tabular}
\caption{Voting thresholds for proposal approval across various cryptocurrencies.}
\label{tab:voting_thresholds}
\end{table}

\subsection{Security}
All selected cryptocurrencies reveal the votes publicly, therefore none satisfy notions of ballot secrecy. Conversely, verifiability is obviously satisfied, since all votes are public, everyone can check the election outcome is correct. The public nature of votes however, makes coercion resistance impossible. Because a voter can prove how they voted, they can be bribed or threatened to vote a certain way. 
On-chain voting means miners or validators are responsible for processing transactions containing votes. They have the power to exclude transactions from the DLT containing votes, and therefore formal notions of censorship resistance cannot be provably guaranteed. In practise, most miners or validators do include valid transactions and follow the protocol, but censorship resistance is not guaranteed: researchers have found miners may still misbehave by reordering, inserting, or censoring transactions for their own gain \cite{MEV}.

\subsection{Missing properties}
From the evaluation of the selected cryptocurrencies, we find there are a number of currently disregarded good governance properties. These include:

\noindent\textbf{1. Binding Elections:} some governance protocols carry out their decision-making process on off-chain platforms \cite{ethereum_governance, bitcoin_governance}, meaning the results are not binding. Similarly, some schemes allow participants to vote that are not responsible for implementing the results \cite{bitcoin_governance}. 
\newline
\noindent\textbf{2. Censorship resistance:} it must not be possible to suppress voter's votes.
\newline
\noindent\textbf{3. Incentive alignment:} the governance protocol must not create active incentives to attack it. In other words, if voters act following a strategy that maximises their benefit, this strategy should not be one that actively undermines fairness, security and decentralisation in the governance protocol. 
\newline
\noindent\textbf{4. Non-purchasable voting power:} voting rights should not be distributed proportionally to wealth. 
\newline
\noindent\textbf{5. Fee-less voting:} voters should not be required to pay to vote.
\newline
\noindent\textbf{6. Ballot Secrecy:} votes must remain secret. Ballot secrecy should not come at the expense of verifiability.\newline 

 In current DLTs, achieving all the above properties is not possible, since some are mutually exclusive. We encourage practitioners to consider the context specific security and performance needs of their DLT and select a compromise appropriately. We proceed by outlining which properties are mutually exclusive and why that is the case. 

\noindent\textbf{Binding elections and fee-less voting:} to ensure that election outcomes must be implemented on the DLT, these must be held on-chain. However, on-chain voting sacrifices fee-less votes, because to cast a vote on-chain one must pay transaction fees. On-chain voting in DLTs also currently sacrifices ballot secrecy in favour of verifiability, although this need not be the case. 
\newline
\noindent\textbf{Binding elections and censorship resistance:} on-chain voting cannot guarantee censorship resistance. However, this concern is application specific, as some DLTs may offer increased levels of censorship resistance due to their protocol design and network distribution. This is also dependant on how strong a notion of censorship resistance DLT practitioners wish to provide.
\newline
\noindent\textbf{Non-purchasable voting power and permissionless access to the DLT:} this is currently the most prevalent issue in DLT governance. DLTs were designed to avoid having to trust central authorities in order to deliver a service. In order to achieve this whilst being protected against resource attacks, they must still offer some protection against Sybil attacks\footnote{A Sybil attack is one where an agent easily obtains multiple identities in the network to gain unfair advantages \cite{mohaisen2013sybil}.}. 
In other words: because DLTs aim to be decentralised, they must avoid scenarios where any actor can access the network and attack it. Therefore, access to the DLT must be regulated.
Currently, to avoid doing this by having to again rely on a centralised trust authority to grant access credentials, access is regulated through a proxy for wealth. The wealthier amass more voting power as a consequence. De-facto, this re-introduces centralisation in the DLT. We advocate for access control mechanisms that do not prioritise the wealthier, but are based on other criteria. An example is the Tree-Proof-of-Position mechanism presented in \cite{tpop-cdc}. This work presents a decentralised class of algorithms that allow agents to prove they are in a given position in a collaborative manner. A scheme like this may be used to regulate access to DLTs, such that only agents with a valid proof of position may be granted participation rights.
\newline
\noindent\textbf{Secret ballots and verifiability:} current implementations of DLT governance also forgo ballot secrecy in favour of verifiability. Votes are cast publicly on-chain, or are publicly decrypted at the end of an election. A number of e-voting schemes have been developed to achieve both. Examples include: \cite{adida2008helios}, \cite{ryan2009pret}, \cite{clarkson2008civitas}, and even proposals that address coercion resistance such as \cite{juels2005coercion}.

\section{Conclusion}\label{conclusion}

This paper underscores the critical role of governance in ensuring the security, fairness, and sustainability of Distributed Ledger Technologies (DLTs). By identifying and categorising vulnerabilities that arise from poor governance, we have shown how these systems, despite their promise of decentralisation and transparency, remain susceptible to exploitation. The outlined threats—ranging from coercion and vote-buying to centralization of power—highlight the urgent need for robust governance protocols that align with established principles of fairness, security, and accountability.

Our key contributions include:
\begin{itemize}
    \item A taxonomy of governance properties that defines the minimum requirements for ensuring good governance in DLTs.
    \item A comprehensive analysis of vulnerabilities stemming from governance deficiencies and their implications for stakeholders.
    \item Guidance on technical solutions, bridging insights from cryptography, social choice theory, and e-voting systems to propose actionable pathways for practitioners to implement good governance properties.
\end{itemize}

This work is necessary because the stakes are high: as DLTs increasingly intersect with public policy, financial systems, and societal infrastructures, the absence of proper governance safeguards risks undermining trust, enabling scams, heists, hacks, centralisation of power and rug pulls. By addressing these vulnerabilities, we not only protect the integrity of DLTs but also uphold their potential to serve as tools for social good.

Future work should focus on:
\begin{enumerate}
    \item Developing scalable technical solutions for implementing governance properties. Particular focus should be placed on alleviating the severity of properties that are mutually exclusive for good governance. A key point of improvement should be developing permissionless access to DLTs that does not enable purchasable voting power, and delivering fee-less participation.
    \item Empirical evaluation frameworks of governance models should be further developed to measure their efficacy in real-world scenarios.
    \item Interdisciplinary collaboration to integrate insights from regulatory, ethical, and socio-technical perspectives into governance design. It is crucial to include clear and tangible definitions of good governance in new policies and regulation. Otherwise, it will be impossible to hold projects liable and prevent undesirable vulnerabilities.
    \item Standardization efforts to define and promote universally accepted frameworks for good governance in DLTs.
\end{enumerate}

We call upon the DLT community—researchers, developers, users, and policymakers—to prioritize governance as a fundamental component of system design. Only through collective effort and innovation can we ensure that DLTs fulfill their potential as equitable and secure platforms for a decentralized future.

\begin{acks}
Aida Manzano Kharman gratefully acknowledges the IOTA Foundation for funding her PhD studies. William Sanders was employed by the IOTA Foundation during the period in which the research for this paper was conducted. The authors would like to thank Professor Robert Shorten and Dr. Pietro Ferraro for their insightful advice and valuable contributions.
\end{acks}

\bibliographystyle{ACM-Reference-Format}
\bibliography{mybibliography}


\end{document}